**Kondo-like zero-bias conductance anomaly in a three-dimensional topological insulator nanowire**

Sungjae Cho*[1,3], Ruidan Zhong[2], John Schneeloch[2], Genda Gu[2], Nadya Mason[§3]

[1] *Department of Physics, Department of Physics, Korea Advanced Institute of Science and Technology, Daejeon 305-701, Republic of Korea*

[2] *Condensed Matter Physics and Materials Science Department, Brookhaven National Laboratory, Upton, NY 11973, USA*

[3] *Department of Physics and Frederick Seitz Materials Research Laboratory, University of Illinois, 104 South Goodwin Avenue, Urbana, Illinois 61801, USA.*

correspondence to S. Cho(*sungjae.cho@kaist.ac.kr) or N. Mason(§nadya@illinois.edu).

Zero-bias anomalies in topological nanowires have recently captured significant attention, as they are possible signatures of Majorana modes. Yet there are many other possible origins of zero-bias peaks in nanowires—for example, weak localization, Andreev bound states, or the Kondo effect. Here, we discuss observations of differential-conductance peaks at zero-bias voltage in non-superconducting electronic transport through a 3D topological insulator $(Bi_{1.33}Sb_{0.67})Se_3$ nanowire. The zero-bias conductance peaks show logarithmic temperature dependence and often linear splitting with magnetic fields, both of which are signatures of the Kondo effect in quantum dots. We characterize the zero-bias peaks and discuss their origin.

# Introduction

The three dimensional topological insulator (3D TI) is a new class of material having metallic surface states inside a bulk band gap[1-3]. The topological surface states are characterized by gapless Dirac dispersions and novel properties such as momentum-spin locking, which were confirmed by angle-resolved photoemission spectroscopy (ARPES)[4-6], scanning tunneling spectroscopy (STS)[7-11] and electrical transport measurements[12-19]. 3D TI nanowires with an insulating bulk, which can be described as a hollow metallic cylinder, have shown Aharonov-Bohm oscillations when a magnetic flux is threaded through the axis[20,21] and Coulomb blockade behavior when connected to metal electrodes through ultrathin TI tunnel barriers[22]. Recently, TI nanowires in the proximity of s-wave superconductors have been predicted to harbor Majorana bound states[23,24], a transport signature of which is a zero-bias tunneling-conductance peak. Similar proximity-coupled topological nanowire systems of InSb[25] and Fe [26] have recently demonstrated Majorana-like zero-bias anomalies, leading to a worldwide effort to better understand the origin and behavior of zero-bias anomalies in topological wires. Zero-bias conductance peaks, not observed yet in 3D TI nanowires, are novel features which may be related to various physical origins such as weak antilocalization, Andreev bound states, and Kondo effect besides Majorana bound states[27-29]. Here we report the first observation of zero-bias anomalies in non-superconducting electronic transport through a 3D TI nanowire contacted by two metal electrodes. We also observed logarithmic temperature dependence and linear splitting with magnetic fields, both of which imply that the zero-bias peaks result from Kondo-like effect.

## Results

We have grown bulk $(Bi_{1.33}Sb_{0.67})Se_3$ crystals and confirmed existence of the surface states inside the bulk band gap by ARPES, as described elsewhere[30]. Using the "scotch tape method", we obtained naturally cleaved topological insulator nanowires on 300 nm $SiO_2$/highly *n*-doped Si substrates. Subsequently, we characterized and chose nanowires having widths ≤ 100 nm and thickness > 12nm by atomic force microscope to avoid unwanted wavefunction hybridizations of top and bottom surfaces[31]. Widths of nanowires were measured again, after all the electrical measurements were done, by scanning electron microscopy (SEM). Immediately after we chose nanowires and identified their locations on a $SiO_2$/Si chip, we spun electron-beam resist (Microchem Corp. PMMA 950 A4) double layer at 4000 rpm to avoid possible contamination or excessive doping of the nanowire surfaces by long exposure to air[32]. Subsequently, we performed electron beam lithography, developed to remove the resist in the regions of source/drain electrodes, and finally deposited Ti/Al (2.5 nm /150 nm) with Au 10 nm capping following a brief ion milling at low power. The devices were wire-bonded and cooled-down in a commercial dilution refrigerator immediately following lift-off in acetone for 1 hour. Figure 1a) is a false-colored SEM image of a completed device, where a small section of the nanowire having dimensions of width ~ 90 nm, length ~ 90 nm, and thickness ~ 13 nm is contacted by a source and a drain electrode.

Here, we report our two-probe differential conductance measurements made between Ti/Al source-drain electrodes driven normal (non-superconducting). To do this, we applied small perpendicular magnetic field of 30 ~ 200 mT, which is above the critical field of the Al electrodes ~ 12 mT (see Supplementary Info.). Fig 1b) shows that as gate voltage decreased, differential conductance ($dI/dV$) at zero-bias voltage decreased due to decrease in density of states, and finally as gate voltages exceeded $V_g$ = - 80V, the conductance reached its minimum implying the chemical potential tuned near the charge neutrality point and therefore minimum density of states. Due to the low electron doping of our Sb-doped crystals and exfoliated devices as reported previously[21,30], we are able to tune the chemical potential effectively below the bottom of the bulk conduction band using a back-gate through $SiO_2$ in our nanowire device. In all range of gate voltages, we observed reproducible conductance fluctuations reminiscent of Coulomb charging effect or phase coherent interference effects such as Fabry-Perot and universal conductance fluctuations.

Fig. 1(c) shows differential conductance spectroscopy, a two-dimensional plot of $G(V_{sd},V_g)$ measured while varying dc source-drain bias-voltages $V_{sd}$ at different gate voltages $V_g$. The transport spectroscopy unexpectedly showed enhanced zero-bias conductance persisting with gate voltages at large negative gate voltages $V_g < -50$V where the background conductance $\leq 5\ e^2/h$. Fig. 1(d) shows differential conductance as a function of bias voltages at $V_g = -53.7$V. We have further performed differential conductance spectroscopy measurements in wider range of gate voltages as shown in Fig. 2. Zero-bias conductance peaks having amplitudes ranging from 0.05 to 0.7 $e^2/h$ were observed in a wide range of gate voltages. Often, zero-bias conductance dips appeared in certain regions of gate voltages as shown in Fig 2a) and 2c). In the rest of this letter, we discuss the possible origin of the observed zero-bias conductance anomaly.

## Discussions

Recently, zero-bias peaks observed in 1D topological superconductors, i.e. a strong spin-orbit coupled semiconducting nanowire in the proximity of superconductors under a parallel magnetic field, captured significant attention. The zero-bias anomaly in those systems have been explained as a signature of Majorana zero modes[23]. There have been similar theoretical proposals on producing Majorana zero modes in topological insulator ($Bi_2Se_3$) nanowires proximity-coupled to s-wave superconductors under a parallel magnetic field[23,24]. However, the absence of superconductivity in a magnetic field higher than the critical field of Al electrodes in our setup excludes the Majorana zero modes among possible explanations for our zero-bias peaks. Due to lack of superconductivity, Andreev states bound to superconductors also does not provide explanations for our observed zero-bias conductance peaks.

To understand the origin of the zero-bias conductance peaks, we have performed the transport measurements at different magnetic fields. Fig. 3 shows two different ways in which peaks evolve with magnetic fields at fixed gate voltages. Peaks having large amplitude (> 0.4 $e^2/h$) measured at $V_g$=-69V split with magnetic fields (Fig. 3(a) and 3(b)). The formation of zero-bias peaks and the splitting of the peaks with a magnetic field are reminiscent of the Kondo effect in quantum dots. A characteristic feature of the Kondo effect in quantum dots is that the zero-bias peak splits with magnetic field at $V = \pm g\, \mu_B B$ (the Zeeman splitting), where $\mu_B$ is the Bohr magneton. Assuming the relation of the zero-bias peaks to Kondo effect, we estimate the Zeeman g-factor of the surface states in our topological insulator $(Bi_{1.33}Sb_{0.67})Se_3$ nanowire is ~ 15 from the blue line in Fig. 3(b), which is roughly half the reported g factor value of the bulk states in $Bi_2Se_3$[33]. On the other hand, peaks of relatively small amplitude (< 0.1 $e^2/h$) observed at $V_g$=-65V switched to conductance dips without splitting as a magnetic field increased (Fig. 3(c)). Collapse of the zero-bias peaks having small peak amplitudes (< 0.1 e2/h) with a magnetic field without being splitted at large background conductance is most likely to be due to lower Kondo temperatures ($T$~$T_K$) at these gate voltages[34].

To further investigate the observed Kondo-like anomaly at zero-bias voltage, we measured temperature dependence of the zero-bias peaks. Fig. 4 shows that conductance of both peaks having small and large amplitudes decreased logarithmically with temperature. Small peaks with large background conductance vanished at relatively low temperatures < 1K as shown in Fig 4(a), and the large peaks with relatively small conductance background did not vanish up to the temperature limit (~ 1.8K) in our dilution refrigerator. Both logarithmic temperature dependence and peak splitting with magnetic field suggest that the zero-bias peaks are Kondo-like effect in a quantum dot. We obtain rough estimates of the Kondo temperatures, $T_{K,}$ ranging from 300 mK to 5 K by equating the full-width at half-maximum of different zero-bias conductance peaks at the base temperature to $2k_B T_K/e$[35,36].

The Kondo effect describes the upturn of the resistance of metals at low temperatures when magnetic impurities are added[37]. A similarKondo effect in semiconducting nanostructures results from a bound state formed between a local spin in a quantum dot and the electrons in the reservoir of source/drain electrodes. We find that the overall background conductance values in our experiments are relatively high ($3e^2/h < G < 5e^2/h$) compared to the conductance values ($G < 2e^2/h$) previously reported in quantum dots showing Kondo effect[35,36], and that our topological insulator nanowire device behaves as an open quantum dot. Coulomb charging is required in order for the Kondo effect to be observed in quantum dots; this is usually observed in a system having low dot-electrode transmission probabilities and conductance less than $2e^2/h$. However, Coulomb charging effects have often been observed in open quantum dots where $2\ e^2/h < G < 6\ e^2/h$ [38-40]. Therefore, high overall conductance values of $3e^2/h < G < 5e^2/h$ does not necessarily excludes the possibility of Coulomb charging and Kondo effect in our open quantum dot device. The large, oval regions of low-conductance shown in the 2D transport spectroscopy $G(V_{sd}, V_g)$ of Fig. 2(a) at $-54 < V_g < -53$, $-52 < V_g < -51$and $-50.5 < V_g < -49.5$ are consistent with diamond-shaped Coulomb blockade in the presence of high lead-dot transparency. Although we do not clearly observe the even-odd parity behavior typical for the Kondo effect in quantum dots, several experiments of Kondo resonances have reported the absence of even-odd parity behavior[41-43], due to either the formation of higher spin states (spin-triplet Kondo resonances) or correlation effects due to electron-electron interaction dominating over the confinement effect in quantum dots.

We often observe zero-bias peaks that increase with increasing magnetic field (see Supplementary Info.). This has previously been explained by singlet-triplet transitions of electron spin states in a quantum dot. [27] The magnetic-field-induced zero-bias peaks persist up to magnetic fields as large as $B$ = 630mT. We currently do not understand the physical mechanism of this magnetic-field-induced zero-bias peak, although it is possible that they are related to the unique spin-momentum locking on the surface of topological insulators. In this case, it may be more energetically favorable to create a multi-particle triplet state than a single-particle spin-1/2 state. To our knowledge, no theoretical and experimental research has been reported on the subject and further studies are required to understand the physics of possible singlet-triplet transitions in topological insulator nanowires. Pikulin *et al.* pointed out in their simulation studies that weak antilocalization by disorder can also be a source of zero-bias conductance peaks at non-zero magnetic fields which break time-reversal symmetry[29]. However, this scenario is only possible when the 1D system has particle-hole symmetry resulting from superconductivity. Without the particle-hole symmetry, weak antilocalization effect should disappear when a magnetic field breaks time-reversal symmetry. The behavior of magnetic-field-induced zero-bias peaks persisting up to $B$ = 630mT in our device cannot be explained by weak antilocalization effect considering the absence of superconductivity. Moreover, lowering background conductance by increasing the tunnel barriers between electrodes and nanowires by a gate voltage is expected to suppress the zero-bias peaks originating from weak antilocalization. Our observation is opposite to this scenario: zero-bias peaks are absent in the gate voltage regions of $V_g \geq$- 40V where conductance is higher, but more prominent as the gate voltage reduces to negative direction below -50V and conductance decreases. This observation implies that weak antilocalization is not the origin of the zero-bias peaks in our device.

## Conclusion

In conclusion, we have observed zero-bias conductance peaks in non-superconducting transport through a topological insulator nanowire contacted by source-drain electrodes. The logarithmic temperature dependence and splitting of the peaks with magnetic fields strongly imply that the zero-bias peaks occur from Kondo-like origin in a quantum dot. Additional features different from typical Kondo effect in quantum dots such as high background conductance (> 2 $e^2/h$) and absence of even-odd parity behavior were observed, which may be consistent with a singlet-triplet Kondo effect and related to the topological nature of the nanowires.

# Methods

### Device fabrication and measurement

Topological insulator nanowires were obtained by mechanical exfoliation ('scotch tape method') from bulk crystals of $Bi_{1.33}Sb_{0.67}Se_3$, which were grown by a modified floating zone method[27]. After mechanical exfoliations of bulk crystals onto 300nm $SiO_2$/highly *n*-doped Si substrates, the nanowires were found under optical microscope. The dimensions of nanowires were measured by Atomic Force Microscopy and Scanning Electron Microscopy. Electron beam lithography and metal (Ti/Al/Au=2.5nm/150nm/10nm) deposition were used to pattern two-terminal devices on the nanowires. Completed devices were wire-bonded and cooled-down in a commercial dilution refrigerator (base temperature = 16mK). The electrical measurements were performed using standard ac lock-in techniques.

# Acknowledgements

N.M. and S.C. acknowledge support from the ONR under grant N0014-11-1-0728 and N00014-14-1-0338. S.C acknowledges support from the National Research Foundation of Korea(NRF) grant funded by the Korea government(MSIP) (grant no. 2011-0030046). Device fabrication was carried out in the MRL Central Facilities (partially supported by the DOE under DE-FG02-07ER46453 and DE-FG02-07ER46471). The work at BNL was supported by the US Department of Energy, Office of Basic Energy Sciences, under contract DE-AC02-98CH10886. S.C. acknowledges useful discussions with H. Sim and E.G. Moon.

# Contributions

S.C. fabricated the devices and performed the electrical measurements. R.Z., J.S. and G.G. grew the bulk TI crystal. S.C. and N.M. analyzed the data and wrote the paper.

# Competing financial interests

The authors declare no competing financial interests.

**Figure Captions**

**Figure 1| Device image and characterization. (a)** Fault-colored Scanning Electron Microscope image of the two-terminal topological insulator nanowire device. **(b)** Two-probe differential conductance d$I$/d$V$ as a function of back-gate voltage $V_g$ at $B$ = 50mT and $T$ = 16 mK at zero bias-voltage. **(c)(d)** Two-dimensional plots of G($V_{sd}$,$V_g$) and G($V_{sd}$) at $V_g$=-53.7V measured at perpendicular mangeitc field B=200mT.

**Figure 2| Differential conductance spectroscopy. (a)(d)(f)** Two-dimensional plots of G($V_{sd}$,$V_g$) measured with perpendicular mangeitc field B=200mT applied in different gate voltage ranges. **(b)(c)(e)(g)** Differential conductance as a function of bias voltages at fixed gate voltages, **(b)**-53.3V, **(c)**-53.0V, **(e)**-74.4V, and **(g)**-131.6V. Each of those gate voltages where the differential conductance was plotted are marked with light blue dotted-lines in the two-dimensional plots **(a)(d)(f)**.

**Figure 3| Evolution of zero-bias conductance peaks with magnetic fields. (a)** Differential conductance as a function of bias voltages at a fixed gate $V_g$=-69V and different magnetic fields. **(b)** two-dimensional plots of **(a)**. blue lines show splitting of the zero-bias conductance peaks with magnetic fields. **(c)** Differential conductance as a function of bias voltages at a fixed gate $V_g$=-65V and different magnetic fields.

**Figure 4| Temperature dependence of zero-bias conductance peaks.** Temperature dependence of zero-bias conductance peaks observed at **(a)(b)** Vg=-60V, and **(c)(d)** Vg=-72.7V. Conductance of both peaks having small and large amplitudes decreased logarithmically with temperature.

**Figure 1.**

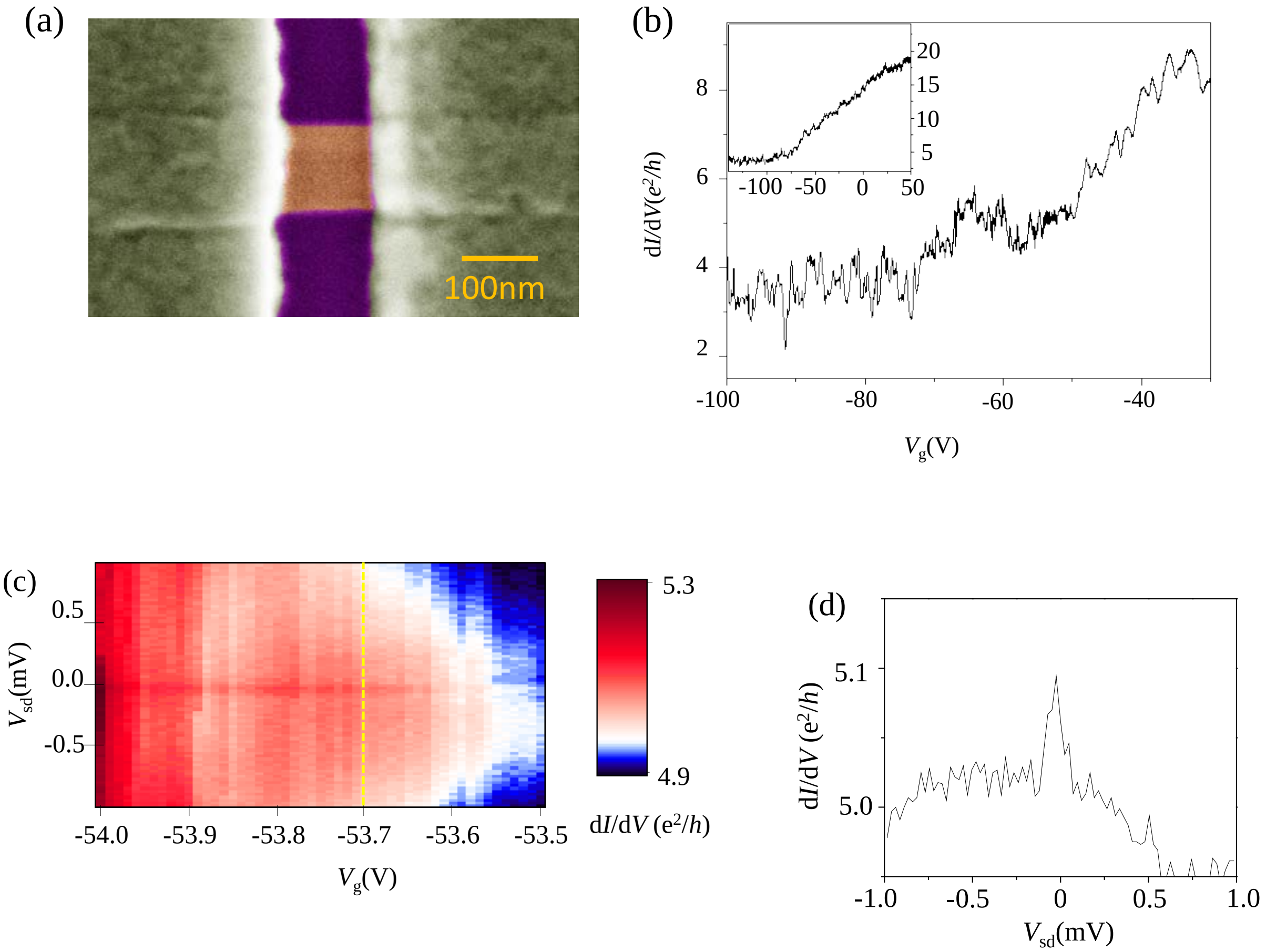

**Figure 2.**

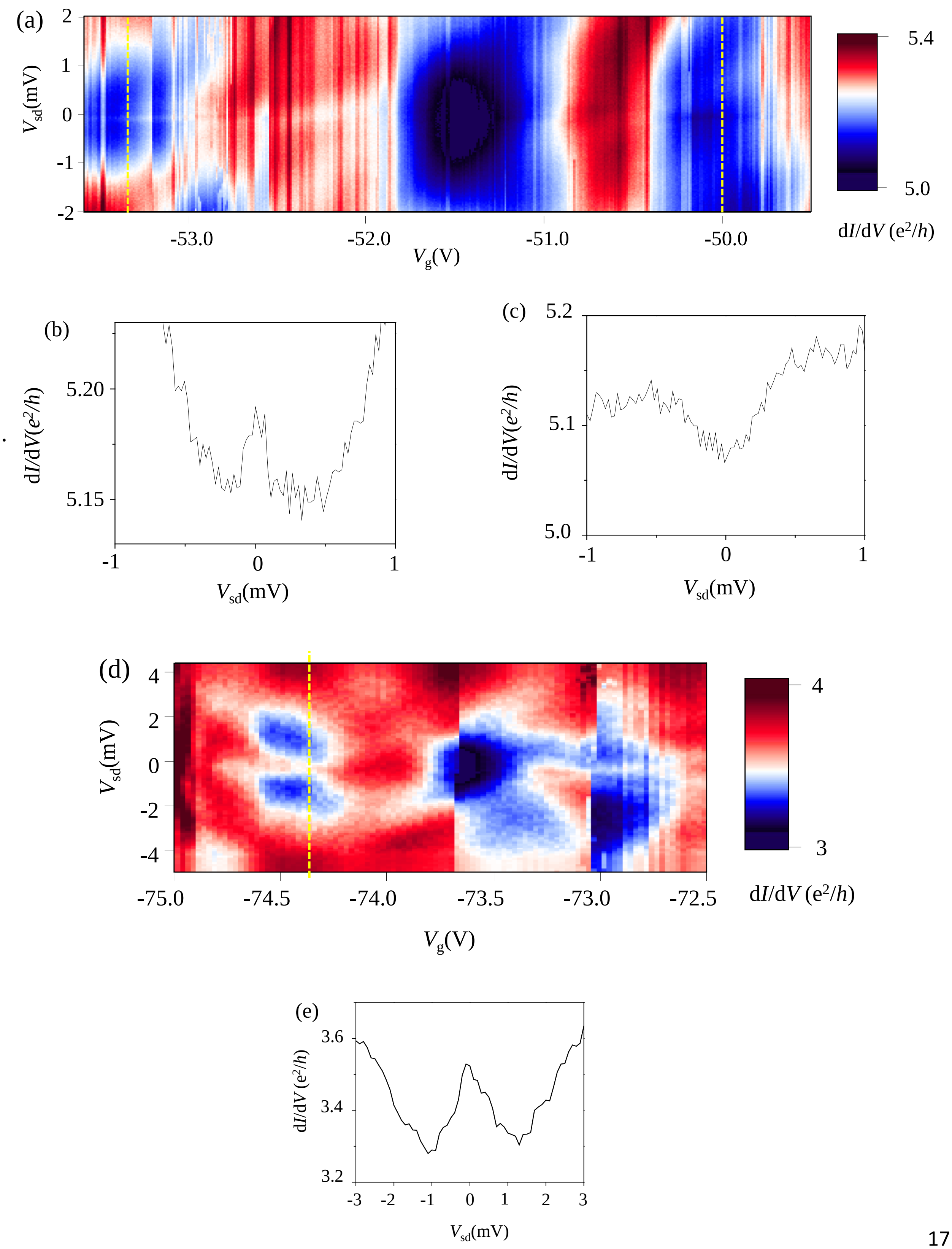

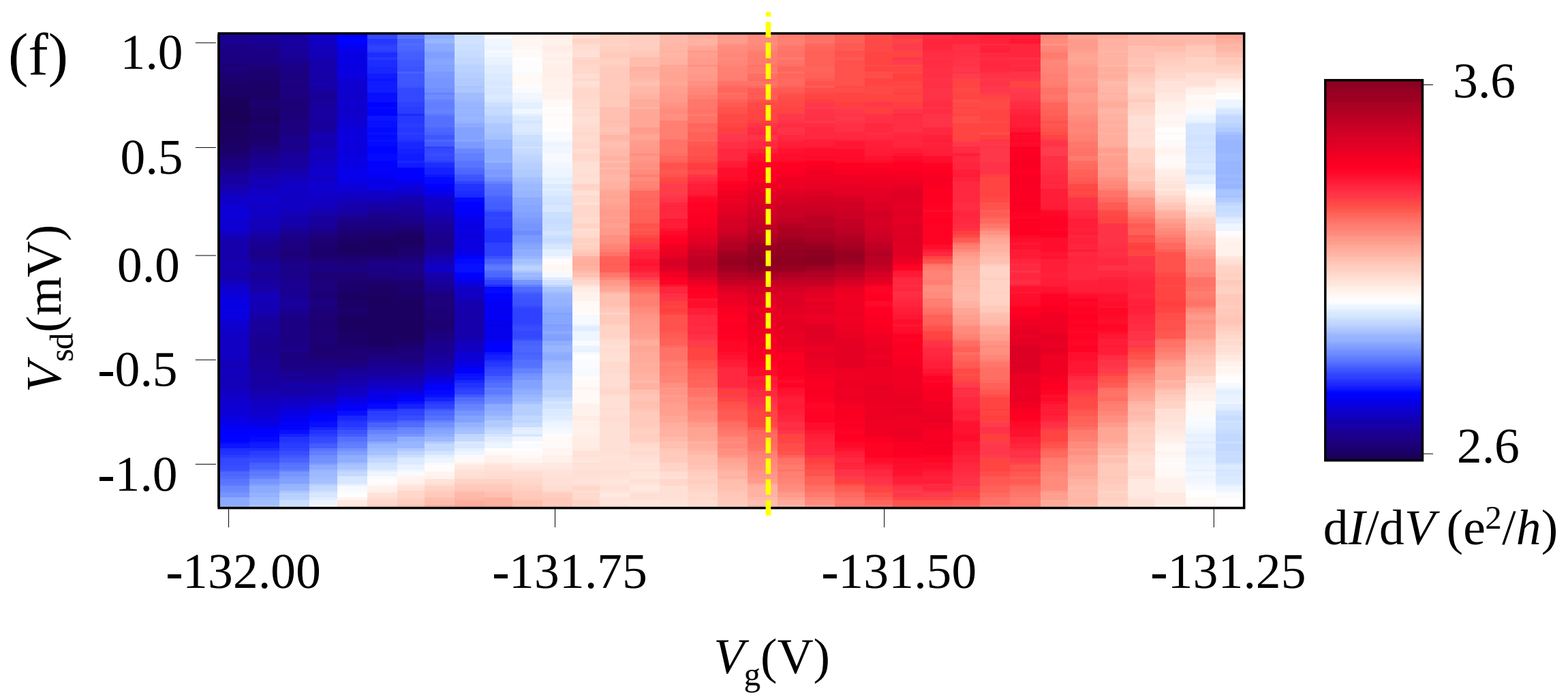

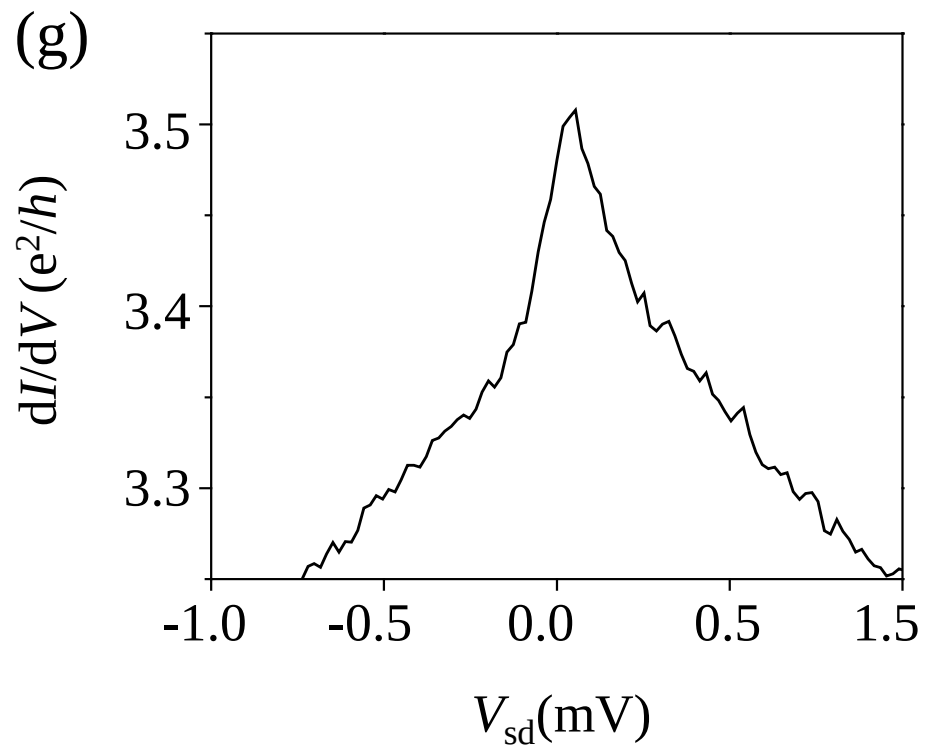

**Figure 3.**

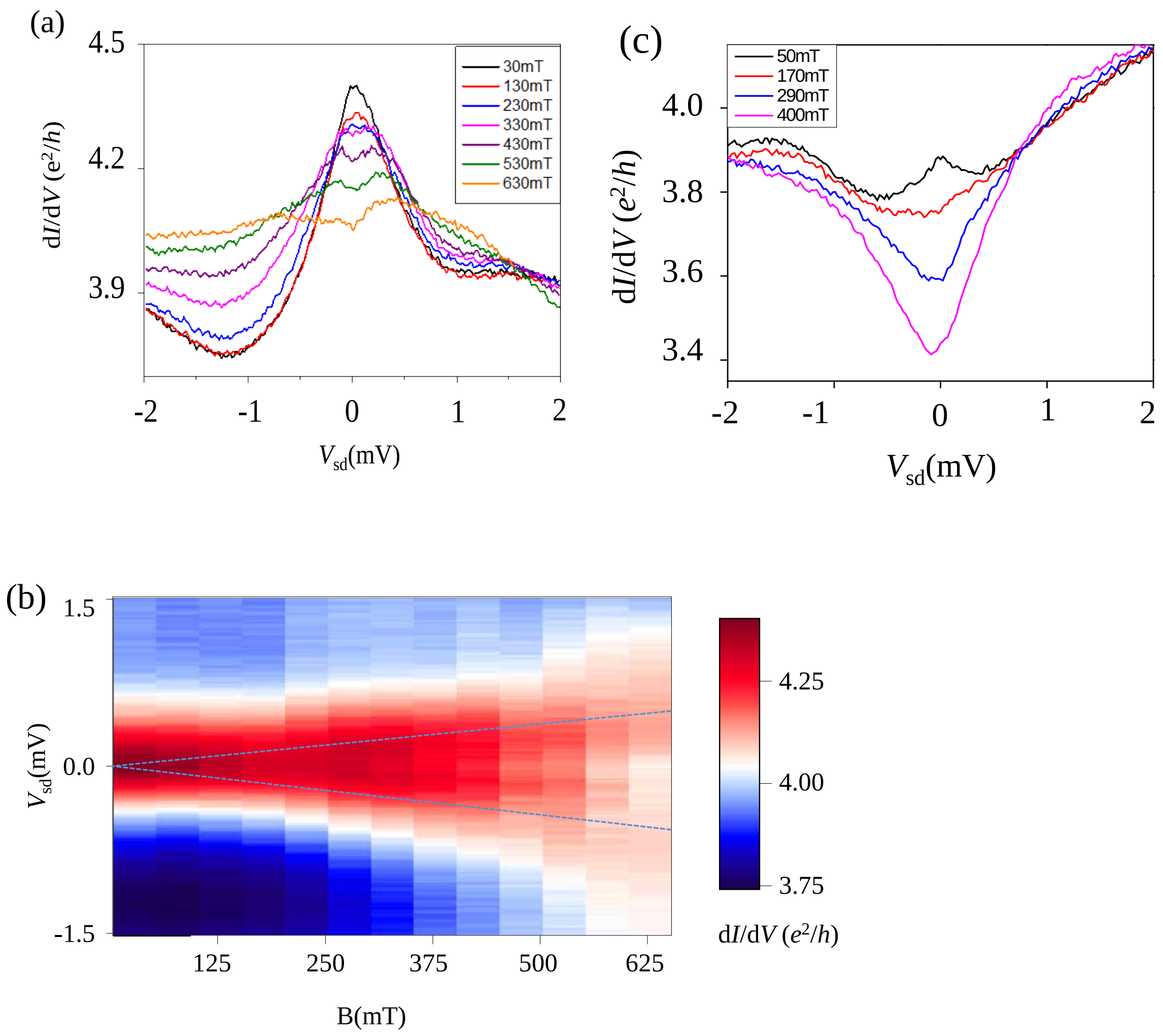

**Figure 4.**

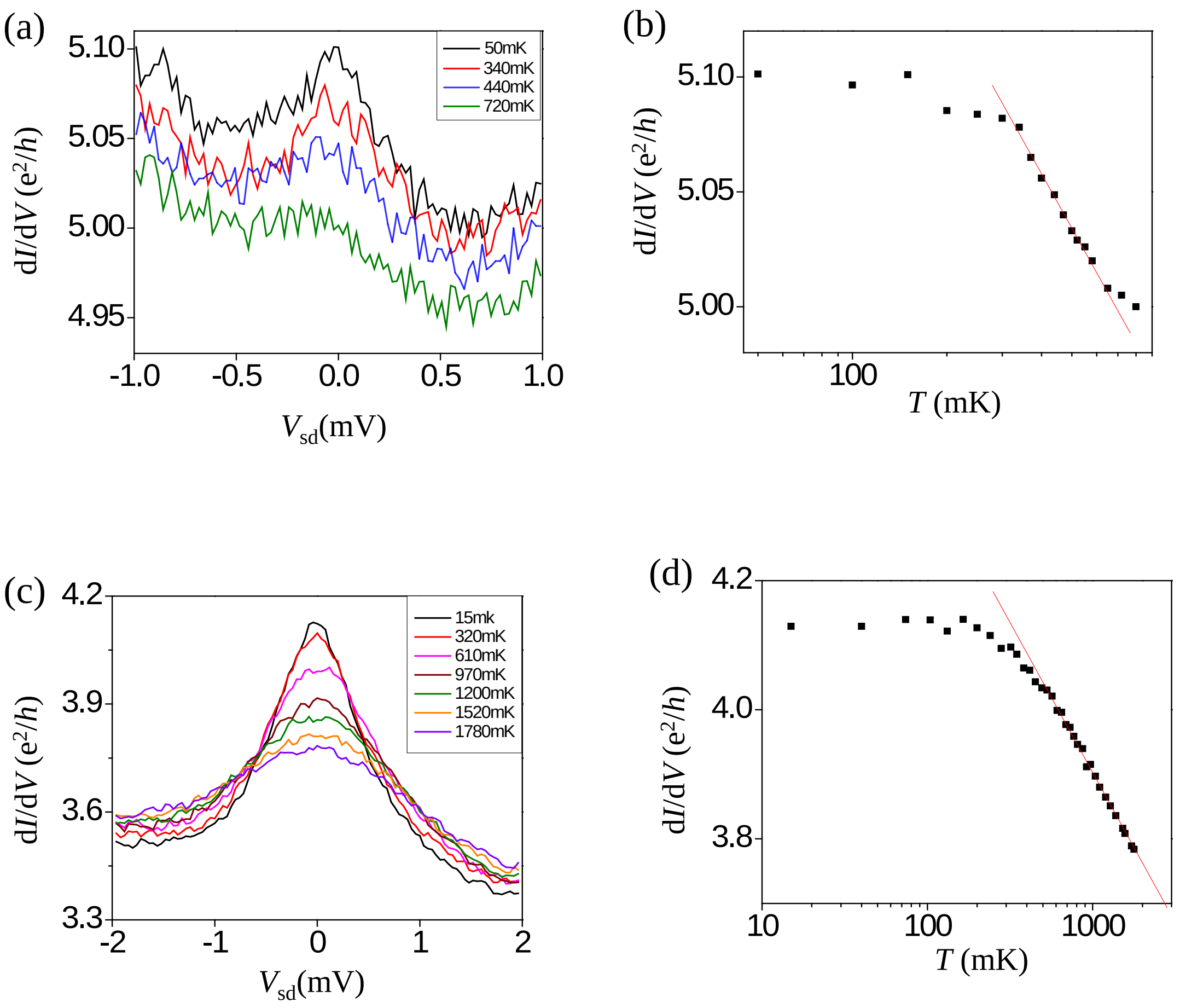

**Kondo-like zero-bias conductance anomaly in a non-superconducting three-dimensional topological insulator nanowire**

Sungjae Cho*[1,3], Ruidan Zhong[2], John Schneeloch[2], Genda Gu[2], Nadya Mason§[3]

[1] *Department of Physics, Department of Physics, Korea Advanced Institute of Science and Technology, Daejeon 305-701, Republic of Korea*

[2]*Condensed Matter Physics and Materials Science Department, Brookhaven National Laboratory, Upton, NY 11973, USA*

[3] *Department of Physics and Frederick Seitz Materials Research Laboratory, 104 South Goodwin Avenue, University of Illinois, Urbana, Illinois 61801, USA.*

correspondence to S. Cho(*sungjae.cho@kaist.ac.kr) or N. Mason(§nadya@illinois.edu).

## Supplementary Discussion

### A. Critical magnetic field of aluminum electrodes

The electrodes of our device consist of Ti(2.5nm)/Al(150nm)/Au(10nm) as described in the Method. We found the critical field value of the electrodes to be ~ 12mT by performing magneto-resistance measurement. The two-probe resistance measurement as a function of perpendicular magnetic field showed sharp transition near B=12mT as shown in Fig. S1 (blue arrows). Below this magnetic field we find that electrodes are superconducting. The measurement reported in the main manuscript was performed with magnetic field applied above this critical field value to ensure that the electrodes are non-superconducting.

### B. Superconducting transport in the topological insulator nanowire device

When a magnetic field is below 12mT, the electrodes are superconducting. Therefore we observed Josephson supercurrent through the topological insulator nanowire in this magnetic field regime. Fig. S2 (a) shows differential resistance (d$V$/d$I$) as a function of gate voltages. Our nanowire device shows finite supercurrents at $V_g$>-50V and the critical supercurrents, the currents at the boundary between the dark blue region (d$V$/d$I$=0) and the region outside the dark blue, increase as density of states increases with the gate voltage. The critical current dependence on gate voltage is very similar to the reported Josephson effect experiment in 3D topological insulator films[1].

**C. Magnetic field dependence of zero-bias conductance peaks**

Fig. S3 shows two-dimensional plots of G($V_{sd}$,$V_g$) at different magnetic fields applied perpendicular to the substrate. A pair of sharp resonances crosses zero energy at Vg = -96.5V and -96.2V. These sharp resonances most likely originate from Fabry-Perot like interference. Fig. S3 shows an interesting magnetic field dependence of the zero-bias conductance anomaly. Some peaks (denoted as blue dotted-lines) do not split with magnetic field, and instead collapse to dips. The amplitudes and widths of these peaks are usually very small and we ascribe the absence of magnetic-field-induced splitting to low Kondo temperature of those peaks. Unexpectedly, we find that often conductance dips at low fields (yellow dotted-lines in Fig. S3) changes into conductance peaks as a magnetic field increases. These magnetic-field-induced peaks persist up to B=630mT. Similar phenomena were observed and explained by singlet-triplet transitions[2]. No theoretical and experimental studies about singlet-triplet transitions in topological insulator quantum dots have been reported and such phenomena are not clearly understood.

**Supplementary Figure Captions**

**S1| Critical magnetic field of Ti/Al/Au electrodes. (a)(b)** Typical magneto-resistance of Ti/Al/Au electrodes in the device as a function of perpendicular magnetic field. **(b)** is a plot of **(a)** in a smaller range of magnetic field.

**S2| superconducting transport in the topological insulator nanowire. (a)**Two-dimensional plots of d*V*/d*I* versus gate voltage $V_g$ and current I measured at B=0. The dark blue regions show the superconducting transport with zero resistance. **(b)** I-V curve cut from (a) along the yellow dotted-line shows typical supercurrent behavior.

**S3| Two-dimensional plots of G($V_{sd}$,$V_g$) at different magnetic fields.** Two-dimensional plots of differential conductance measured in the gate voltage ranges -96.0 V < $V_g$ > -95.5 at four different perpendicular magnetic fields, **(a)** B=30m, **(b)** B=230m**, (c)** B=430m and **(d)** B=630m. Vertical dotted-lines denotes peak(blue) or dip(yellow) locations at zero-bias voltage at B=30mT.

**S1.**

(a)

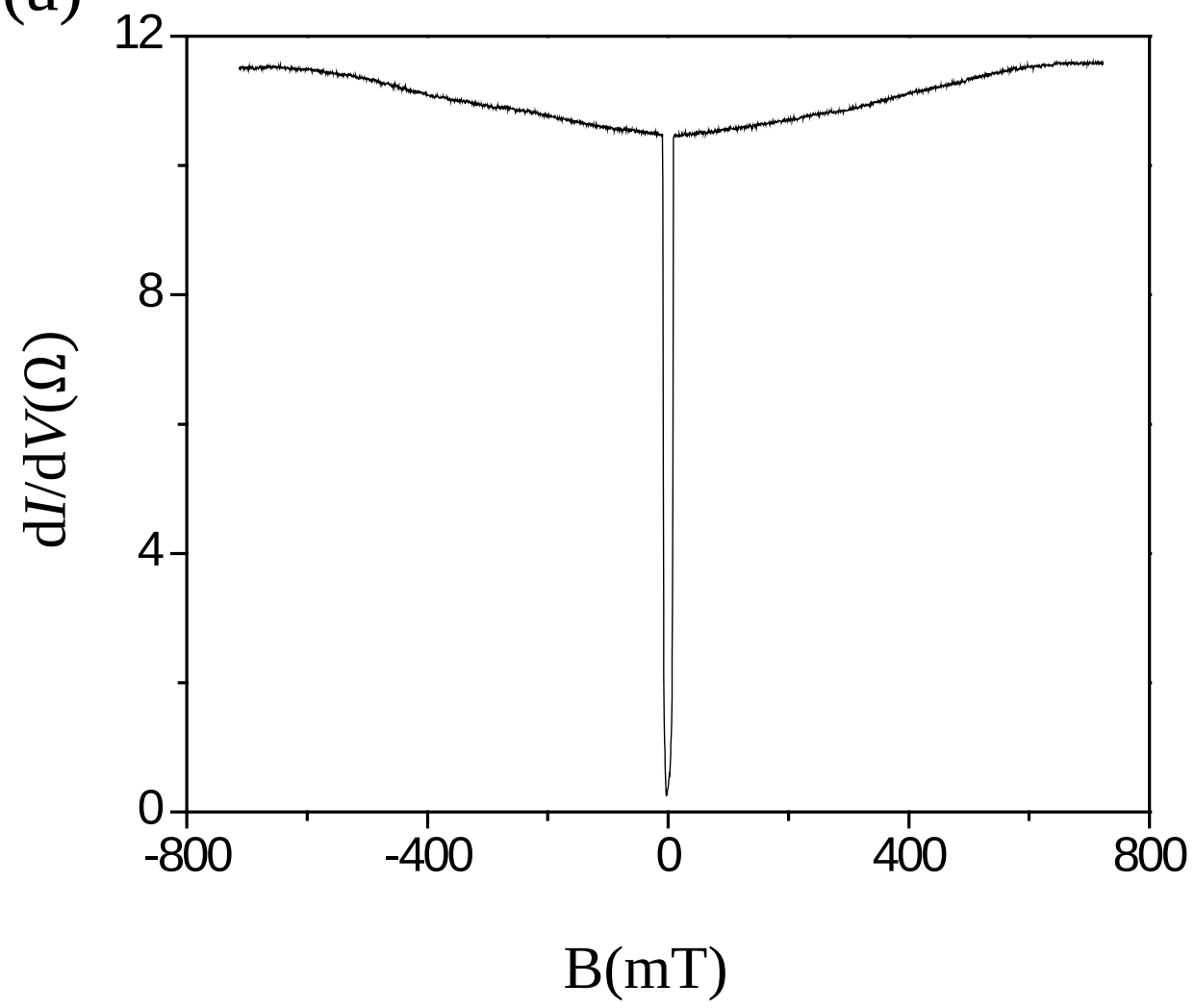

(b)

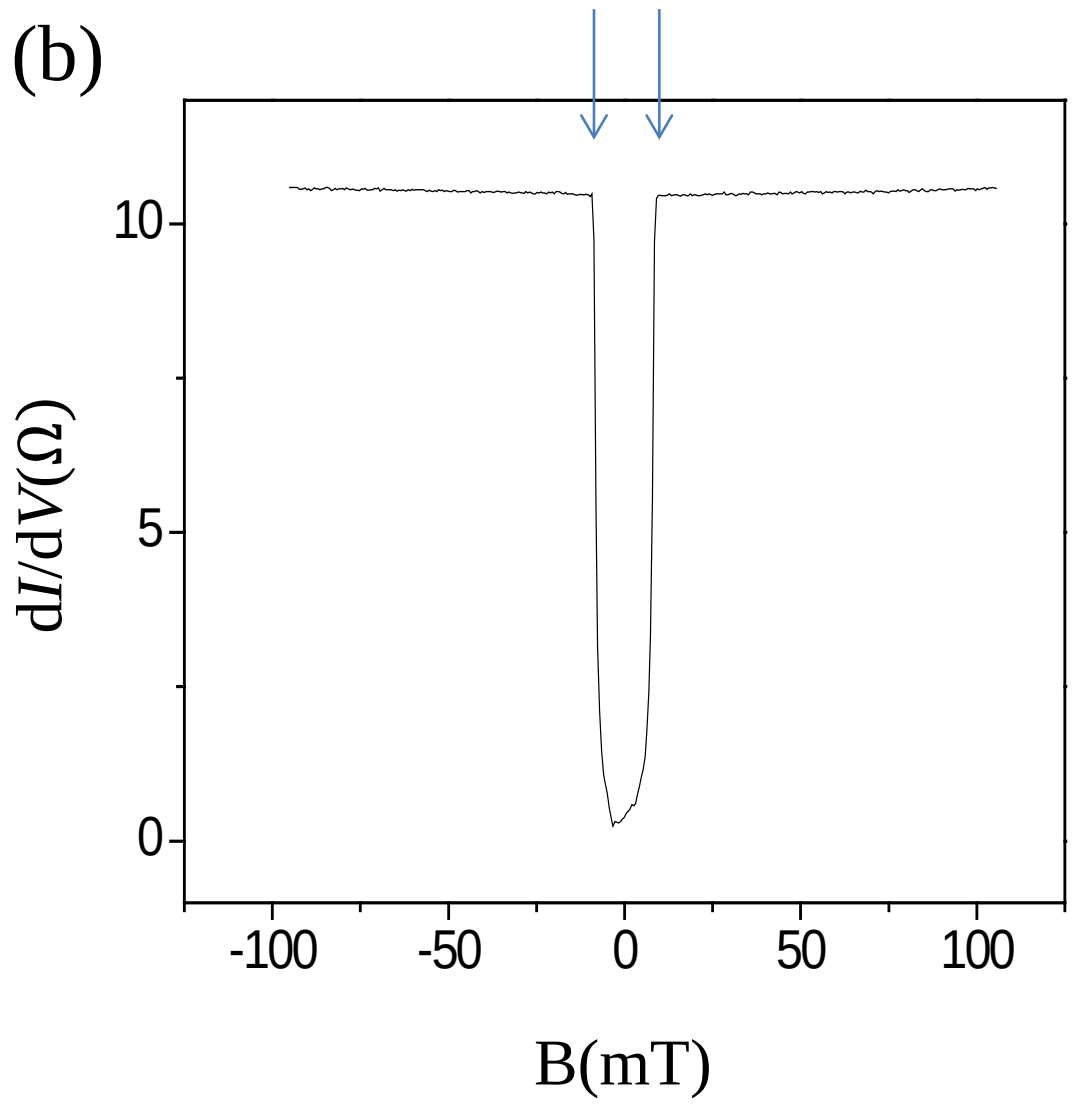

**S2.**

(a)

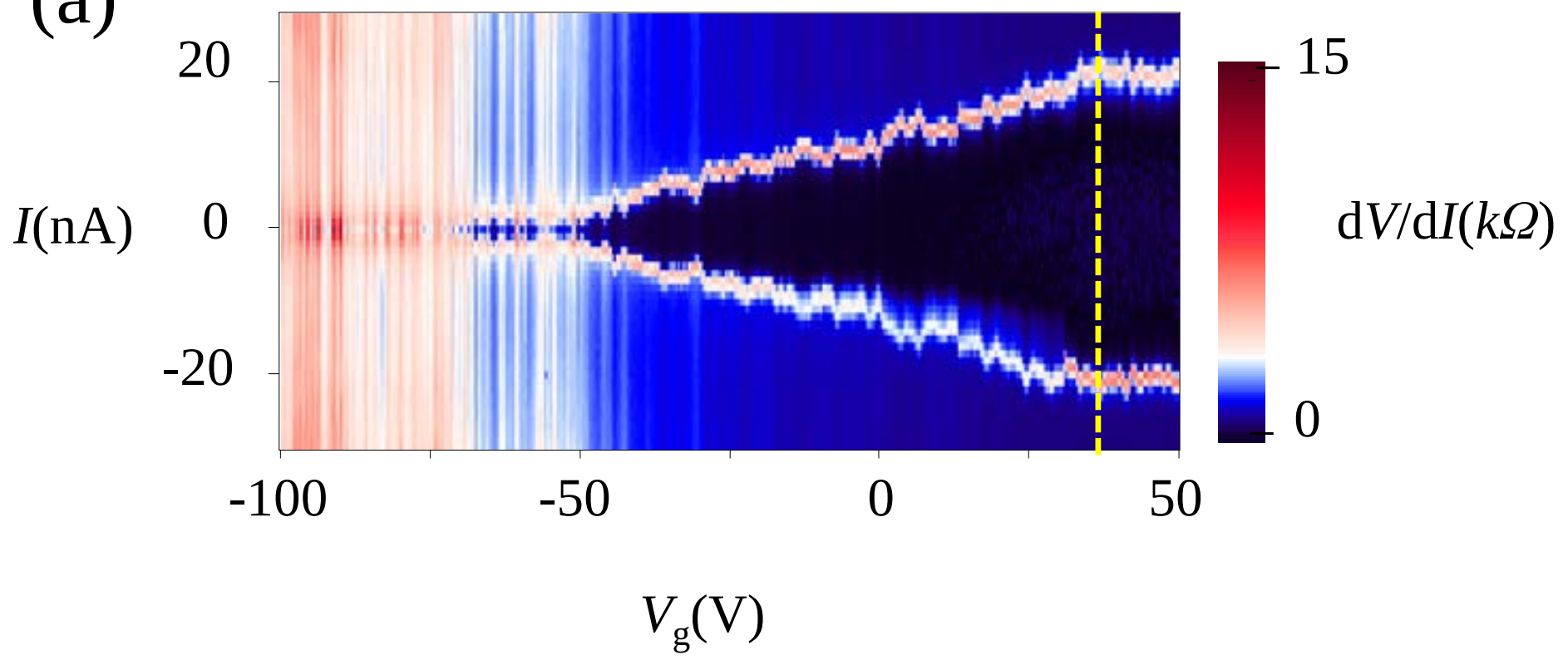

(b)

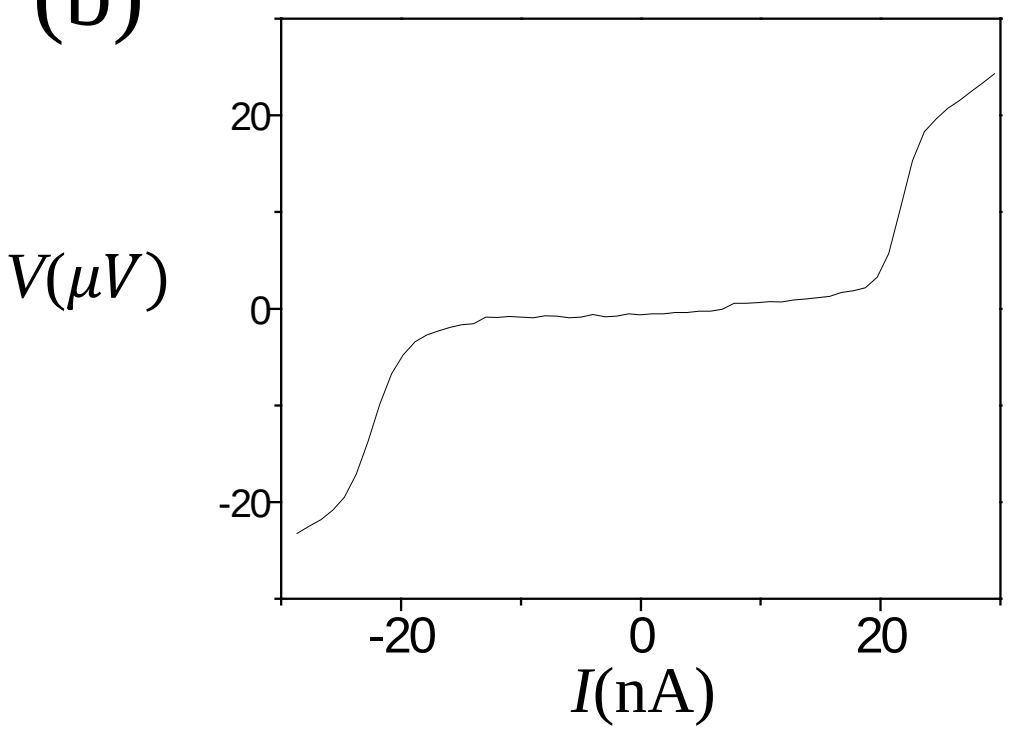

**S3.**

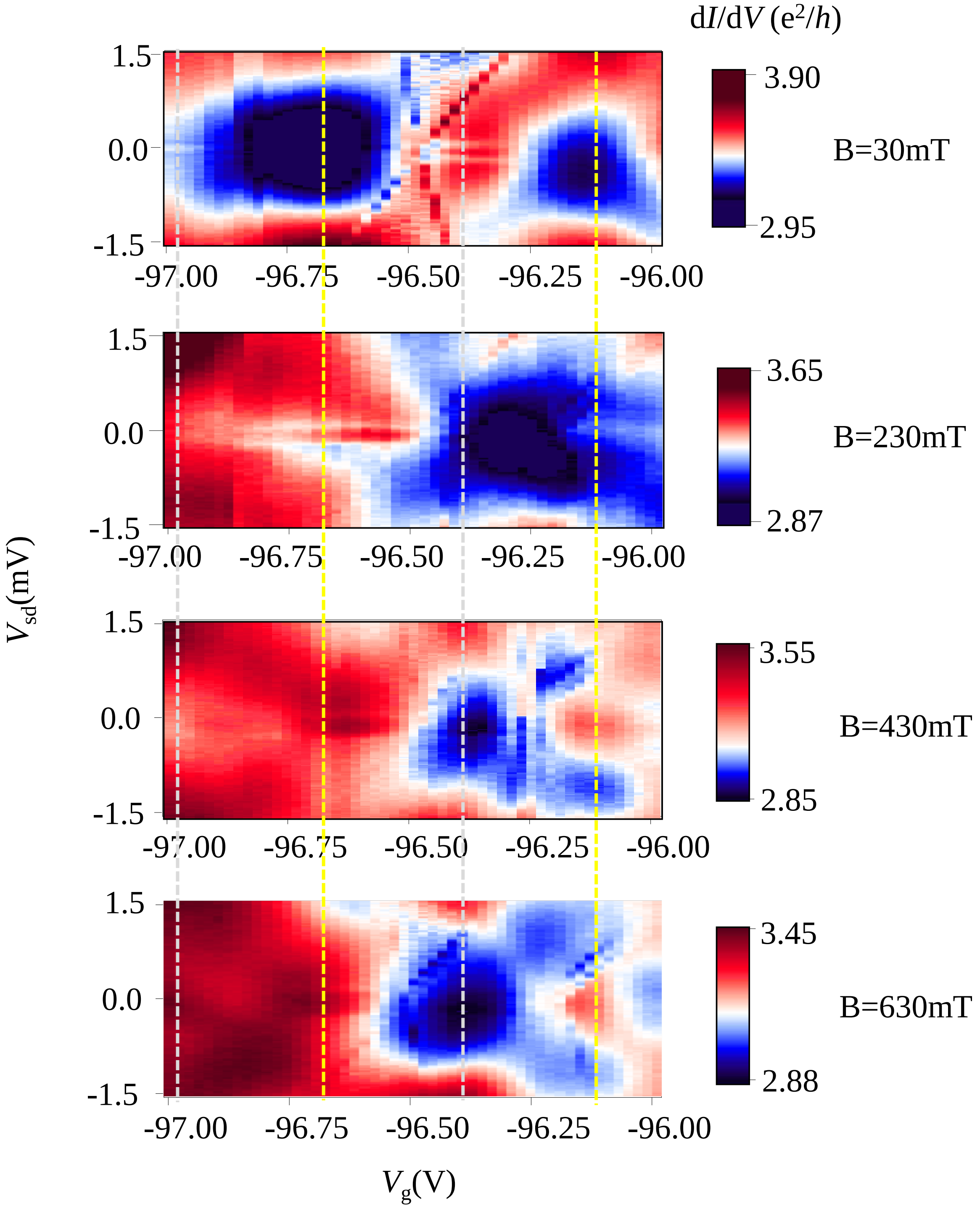